\documentclass{article}
\usepackage{a4,epsf,epsfig,wrapfig}
\newcommand{\A}{{\mathcal A}}
\newcommand{\p}{{\mathcal A}}
\newcommand{\be}{\begin{equation}}
\newcommand{\ee}{\end{equation}}
\newcommand{\ba}{\begin{eqnarray}}
\newcommand{\ea}{\end{eqnarray}}
\newcommand{\bb}{\begin{thebibliography}}
\newcommand{\eb}{\end{thebibliography}}

\newcommand{\lab}[1]{\label{#1}}

\begin{document}
\begin{center}
{\bfseries  INTERACTIONS AT LARGE DISTANCES AND SPIN EFFECTS \\
            IN NUCLEON-NUCLEON AND NUCLEON-NUCLEI SCATTERING\footnote{
To appear in the proceedings of the Xth Workshop on High-Energy Spin Physics,
Dubna, September 16-20, 2003.} }
\vskip 5mm
J.-R. Cudell$^{1}$, E. Predazzi$^{2}$ and
  \underline{O.V. Selyugin}$^{1,3}$
\vskip 5mm
{\small
(1) {\it
 Institut de Physique, B\^at. B5a, Universit\'e de Li\`ege, Sart Tilman, B4000
  Li\`ege, Belgium
}
\\
(2) {\it
Dipartimento di Fisica Teorica - Unversit\`{a} di Torino
   and Sezione INFN di Torino, Italy 
}
\\
(3) {\it
 Bogoliubov Laboratory of Theoretical Physics, JINR, Dubna, Russia.
}
\\
$\dag$ {\it
E-mail: selugin@thsun1.jinr.ru
}}
\end{center}
\vskip 5mm

\begin{abstract}
  The momentum-transfer dependence of the slopes of the spin-non-flip 
 and spin-flip amplitudes is analysed.
  It is shown that the long tail  of the hadronic potential
  in  impact parameter space
%  representation 
   leads for hadron-hadron interactions 
  to a larger value of the slope for the reduced spin-flip
   amplitude than for
 the spin-non-flip amplitude.
  It is shown that the preliminary measurement of $A_N$ obtained by the E950 
Collaboration confirms such a behaviour of the hadron spin-flip amplitude. 
\end{abstract}

\vskip 8mm

%\section{Introduction}

 The diffractive polarised experiments at HERA and RHIC
 allow to study the spin properties of the  quark-pomeron and
 proton-pomeron vertices, 
 and  to search  for a possible  odderon contribution.
      This provides an important test of the
 spin properties of QCD at large distances.
In all of these cases,  pomeron exchange
is expected to contribute to the observed spin effects at some level
 \cite{martini}.

 In the general case,
the  form of the analysing power $A_N$ and
 the position of its maximum 
 depend on the parameters of the elastic scattering
 amplitude $\sigma_{tot}$,  $\rho(s,t)$, on the Coulomb-nucleon interference
 phase  $\varphi_{cn}(s,t)$
 and on the elastic slope $B(s,t)$.
  The Coulomb-hadron phase was calculated
 in the entire diffraction domain taking into account  the form factors
 of the  nucleons \cite{prd-sum}.

   The dependence of the hadron spin-flip amplitude on the momentum transfer
  at small angles is tightly connected with the basic structure of the
  hadrons at large distances. We show that
  the slope of the ``reduced'' hadron spin-flip amplitude
  (the hadron spin-flip amplitude without the kinematic factor $\sqrt{|t|}$)
  can be larger
  than the slope of the hadron spin-non-flip amplitude
  as was observed long ago \cite{predaz}.
 
The first RHIC measurements at $p_L = 22 $~GeV/$c$ \cite{an22}  
in $p ^{12}C$ scattering indicated that
$A_N$ may  change sign already at very small momentum transfer.
Such a behaviour cannot be described by the Coulomb-Nuclear Interference
effect alone, and requires some contribution of the hadron spin-flip amplitude.

The total helicity amplitudes can be written
 as $$ \Phi_i(s,t) = \phi^h_{i}(s,t)
        + \phi_{i}^{em}(t) \exp[i \alpha_{em} \varphi_{cn}(s,t)],$$ 
 where $\phi^h_{i}(s,t)$ comes from the pure strong interaction of hadrons,
  $\phi_{i}^{em}(t)$  from the electromagnetic interaction of hadrons
  ($\alpha_{em}=1/137$ is the electromagnetic constant),
   and
  $\varphi_{cn}(s,t)$ is the electromagnetic-hadron interference phase factor.
  So, to determine the hadron spin-flip amplitude  at small angles, 
  one should take  into account all electromagnetic and 
    interference effects.

  As usual, the slope $B$ of the scattering amplitudes
   is defined  as
   the derivative of the logarithm of the amplitudes with respect to $t$.
     For an exponential dependence on $t$, this  coincides
    with the standard slope of the differential cross sections divided by $2$.
  If we define the forms of the separate  hadron scattering
  amplitude as:
\ba
Im \ \A_{nf}(s,t) &\sim& exp(B_{1}^{+} \ t), \ \ \
  Re \ \A_{nf}(s,t) \sim exp(B_{2}^{+} \ t), \nonumber \\
 Im  \tilde{\A_{sf}}(s,t) & \sim &
  \ exp(B_{1}^{-} \ t), \ \ \
  Re \tilde{\A_{sf}}(s,t) \  
\sim  \ exp(B_{2}^{-} \ t),
\ea
 ($\A_{nf}(s,t)$ and $\tilde{\A_{sf}}(s,t)$ are non-flip and ``reduced'' 
 spin-flip amplitudes respectively), 
 then, at small $t$ ($\sim 0 - 0.1 \ GeV^2$), most
 phenomenological analyses assume
$ B_{1}^{+} \ \approx \ B_{2}^{+} \ \approx \
    B_{1}^{-} \ \approx \ B_{2}^{-} . $
    Actually, we  can take the eikonal representation for the scattering
   amplitude
\ba
\phi^{h}_{1}(s,t) & =& - i p \int_{0}^{\infty} \ \rho \ d\rho
 \ J_{0}(\rho q)
 [e^{\chi_{0}(s,\rho)} \  - \ 1 ], \nonumber \\
\phi^{h}_{5}(s,t) &=& - i p \int_{0}^{\infty} \ \rho \ d\rho
 \ J_{1}(\rho q) \ \chi_{1}(s,\rho) \ e^{\chi_{0}(s,\rho)} .
\ea
where $q=\sqrt{-t}$ and 
  $\chi_{0}(s,\rho)$ represents the corresponding interaction potential
  $V_i(\rho,z)$ in  impact parameter space.
  If the potentials $V_{0}$ and $V_{1}$ are
   assumed to have a Gaussian form
  in the first Born approximation, 
  $\phi^{h}_{1}$ and ${\phi_h}^{5}$
   will have  the same Gaussian form

\ba
 \phi^{h}_{1}(s,t) & \sim &  \int_{0}^{\infty} \ \rho \ d\rho
 \ J_{0}(\rho q) \ e^{- \rho^{2}/2R^{2} } \ = \ R^2 \  e^{R^{2} t/2}, 
                                \nonumber \\ 
 \phi^{h}_{5}(s,t) & \sim & \int_{0}^{\infty} \ \rho^2 \ d\rho
 \ J_{1}(\rho q) \  
   \ e^{-\rho^2/(2R^2) } \ = \ q \ \ R^4 \ e^{ R^{2} t/2}  . \lab{f5a}
\ea
  In this special case,
the slopes of
 the  spin-flip and of the  ``residual''spin-non-flip amplitudes are
  indeed the same.

  However, a Gaussian form for  the potential
 is at best adequate to represent   the central part of the
   hadronic  interaction. This form cuts off the Bessel function
  and the contributions at large distances.
   If we keep only the first two terms in a small $x$ expansion 
  of the $J_{i}$, 
 \ba
 J_{0}(x) \ \simeq \ 1 \ - \ (x/2)^2 ; \ \ \ {\rm and} \ \ \
  2 \ J_{1}/x \ = \ (1 \ - 0.5 \ (x/2)^2) , \lab{sbess}
\ea
 the corresponding integrals 
   have the same behaviour  in $q^2$ \cite{tur1}.  
   So, the integral representation   for spin-flip and spin-non flip amplitudes
 will be the same as in (\ref{f5a}).
  If, however, the potential (or the corresponding eikonal)
 has a long tail (exponential or power)
  in  impact parameter,  
  then the approximation (\ref{sbess}) for the Bessel functions
  does not lead to a correct result and one has to perform
 the full integration.

  Let us examine the contribution of  the large distances.
  The Hankel  asymptotic of the Bessel functions at large distances
%  \cite{sprav} 
 are
\ba
 J_{\nu}(z) \ & = & \  \sqrt{2/ \pi z} \ [ P(\nu , z) \ \cos{\chi(\nu,z)} \ -
 \  Q(\nu,z) \  \sin{\chi(\nu, z)} ],  \nonumber \\
    P(\nu,z) \  & \sim & \ \sum_{k=0}^{\infty} \  (-1)^{k} \  
   \frac{(\nu, \ 2k)}{(2z)^{2k} } ,  \ \ \
  Q(\nu,z) \  \sim  \ \sum_{k=0}^{\infty} \ (-1)^{k} \  
          \frac{(\nu, \ 2k+1)}{(2z)^{2k+1} } 
%      \frac{[(4k-1)!!] \ [4k+1)!!]}{(2k!) \ (8z)^{2k+1}};  
    \lab{asbes} 
\ea
 with $ \chi(\nu,z) =  z- (\nu/2 +1/4)$ and  
 $ P_{0}(x)$ and  $Q_{i}(x)$ some polynomials of $x$.
  The leading behaviour at large  $x$  
 will thus  be proportional  $1/\sqrt{q \rho}$.

\newpage Let us calculate the corresponding  integrals in the case 
  of large distances
\ba
\phi^{h}_{1}(s,t) &\sim& { 1\over q^2} \int_{0}^{\infty} \sqrt{x} \left[\left(
1- {0.125\over x}\right) \cos{x} +\left(1+{0.125\over x}\right) \sin{x}\right]  e^{-{x^2\over 2 R^2 q^2}} dx\nonumber\\ 
 &\approx& {R\over q} \  _{1}F_{1}(3/4,1/2,-q^2 R^2/2),\end{eqnarray}
\ba
\frac{\phi^{h}_{5}(s,t)}{q}  
&   \sim&  \frac{1}{q^4}  \int_{0}^{\infty}  \ x^{3/2}  
   \left[\left({0.375\over x}-1\right) \cos{x}+\left(1- \frac{0.375}{x}\right) \sin{x}\right]  
                   e^{-{x^2\over 2 R^2 q^2}} \ dx   \nonumber \\    
 &\approx &{R^{3/2}\over q^{5/2}}  \ _{1}F_{1}(3/4,1/2,-q^2 R^2/2), 
\label{seven}\ea

%%%%%%%%%%%%%%%%%%%%%%%%%%%%%%%%%%%%%%%%%%%%%%%%%%
\begin{wrapfigure}{R}{8cm}
\mbox{\epsfig{figure=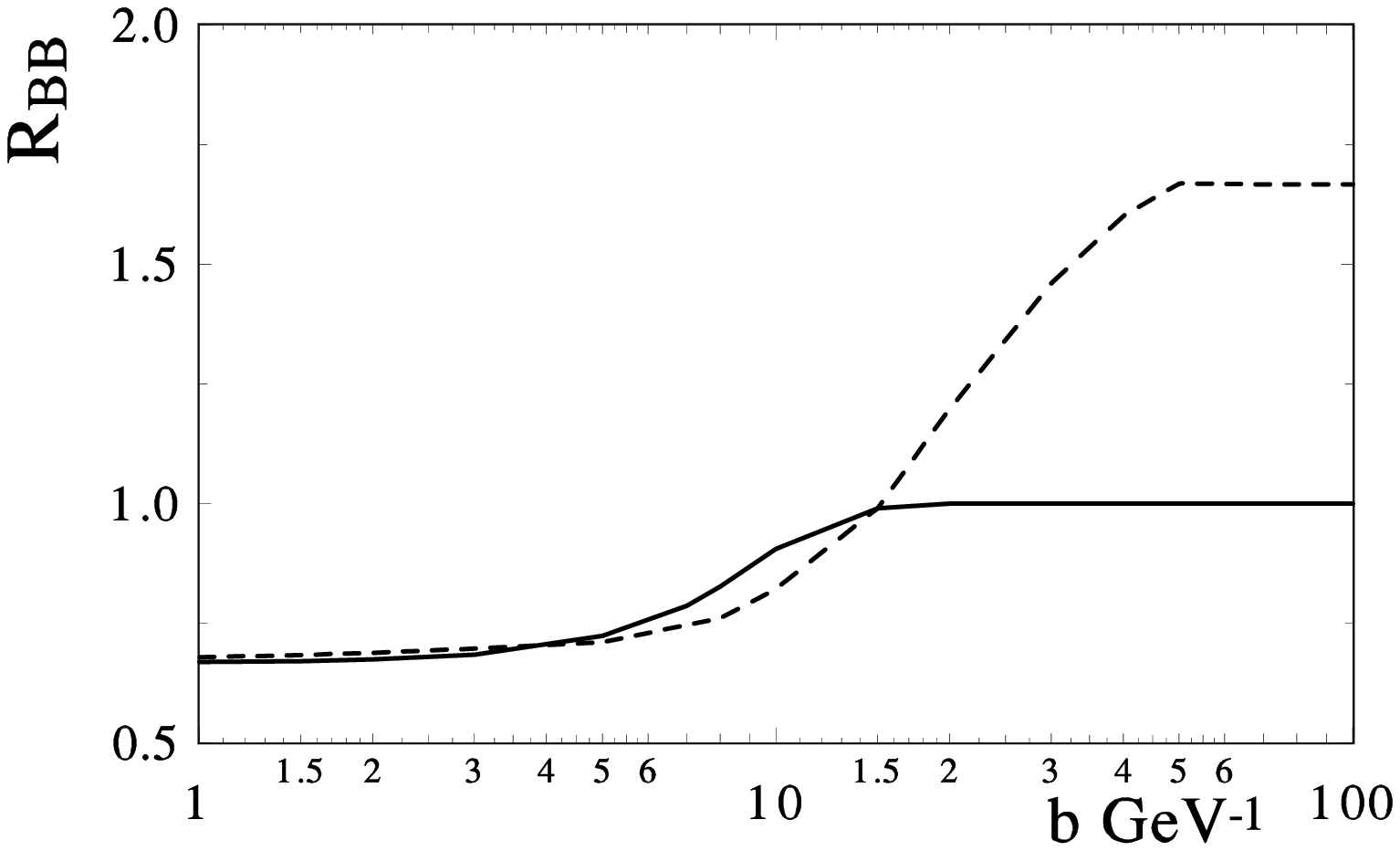,width=7.8cm,height=6cm}}
{\small{\bf Figure 1.}
The ratio of the effective slopes - $R_{BB}$ for the case $n=1$
 (dashed line) and for $n=2$ (solid line) as function of
  the upper bound of the integrals  $b$.
}
%\medskip
\end{wrapfigure}
%To insert figure (with the help of wrapfig.sty)\\

%%%%%%%%%%%%%%%%%%%%%%%%%%%%%

  The exponential asymptotics of  both representations are the same, 
 but the additional $q^{3/2}$ in the denominator of (\ref{seven})
leads to a larger
  slope for the residual spin-flip-amplitude. 
  So, although  the integrals have the same exponential
   behaviour asymptotically, the
  additional inverse power of $q$ leads to  
   a larger  effective slope for the 
  residual spin-flip amplitude at small $q$ 
  although we take a Gaussian representation
  in impact parameter. 
 
  These  investigations are  confirmed numerically.  
    We calculate the scattering amplitude in the Born 
  approximation in the cases
  of exponential and Gaussian form factors in impact parameter space
   as a function of the upper limit $b$ of the corresponding integral
\ba
\phi^{h}_{1}(t) &\sim&  \int_{0}^{b} \ \rho \ d\rho
 \ J_{0}(\rho \Delta) f_{n} , \ \ 
%   \nonumber \\
\phi^{h}_{5}(t)/q \sim  \int_{0}^{b} \ \rho^2 \ d\rho
 \ J_{1}(\rho \Delta) \  f_{n}  . \lab{f5ab}
\ea
 with $f_{n}= \exp{[-(\rho/5)^n]}$, and $n=1,2$.
  We then  calculate the ratio of the slopes of these two amplitudes 
   $R_{BB} =B^{sf}/B^{nf}$ as a function of  $b$ for these two values of $n$.
  The result is shown in Fig.1.
  We  see that at small impact parameter the value of $R_{BB}$ is 
  practically the same in both cases and depends weakly on the value of  
  $b$. But at large distances, the behaviour of $R_{BB}$ is different.
  In the case of the Gaussian form factor,  the value of $R_{BB}$
  reaches its asymptotic value ($=1$) quickly.
   But in the case of the exponential behaviour,
  the value $R_{BB}$ reaches  its limit $R_{BB}=1.7$ only at large distances.
  These calculations confirm our analytical analysis of the asymptotic
  behaviour of these integrals at large distances.

 In \cite{tur1}, it was shown that in the case of an exponential tail
  for the  potentials, 
$ \chi_{i}(b,s) \ \sim \ H \ e^{- a \ \rho}, $
 one  obtains
%{\Large
\ba
 \A_{nf} (s,t) & \sim &  \frac{1}{a\sqrt{a^2+q^2}} \ e^{-B q^2}, \ \  
%\nonumber \\
\sqrt{|t|} \tilde{\A_{sf}}(s,t) \sim
 \ \frac{3 \ a  q \ B^{2}}{ \sqrt{a^2+q^2}} \ \ e^{-2 \ B q^2}.
\ea     
  In this case, therefore,
  the slope of the ``residual'' spin-flip amplitude exceeds the slope
 of the spin-non-flip amplitude by a factor of two.
  Hence, a  long-tail hadron potential
 implies a significant difference in the
  slopes of the ``residual'' spin-flip and of the spin-non-flip amplitudes.

%\section{The analyzing power in the proton-nuclei scattering}

   Recently there has been very few experimental data
   for  hadron-hadron scattering  at large energy. 
  Of course, it will be very interesting to obtain  data from the  
  $PP2PP$ experiment at RHIC. 
  But now, we only have  the preliminary data
  of $A_N$ in  proton-Carbon elastic scattering. Despite the fact
  that these data
   have bad normalisation conditions, the slope of the analysing power is
   very interesting.

    In our analysis, the scattering amplitude is
$
 \A_i(s,t) = \p^h_{i}(s,t) 
        + \p_{i}^{em}(t) e^{i\delta}, (i=nf, sf),
$
where each term includes a hadronic and an electromagnetic contribution
with the Coulomb-nuclear phase \cite{prd-sum}.
 The electromagnetic form factor  $F^{^{12}C}_{em}$ was obtained
from the electromagnetic density of the nucleus. 
We parametrise the  spin non-flip and 
  spin-flip part of $p ^{12}C$ scattering as    
\ba
\A^{pA}_{nf}(s,t) &=& (1+\rho^{pA}) 
{\sigma_{tot}^{pA}(s)\over 4\pi} \exp\left({B^{+}\over 2}t\right) \nonumber \\
{\A^{h}_{sf}}(s,t)&=&   (k_2  + i  k_1) 
  { \sqrt{|t|} \ \sigma_{tot}^{pA}(s)\over 4\pi} 
\exp\left({ B^{-}\over 2}t\right).  
\ea

%%%%%%%%%%%%%%%%%%%%%%%%%%%%%%%%%%%%%%%%%%%%%%%%%%
\begin{wrapfigure}{R}{8cm}
\mbox{\epsfig{figure=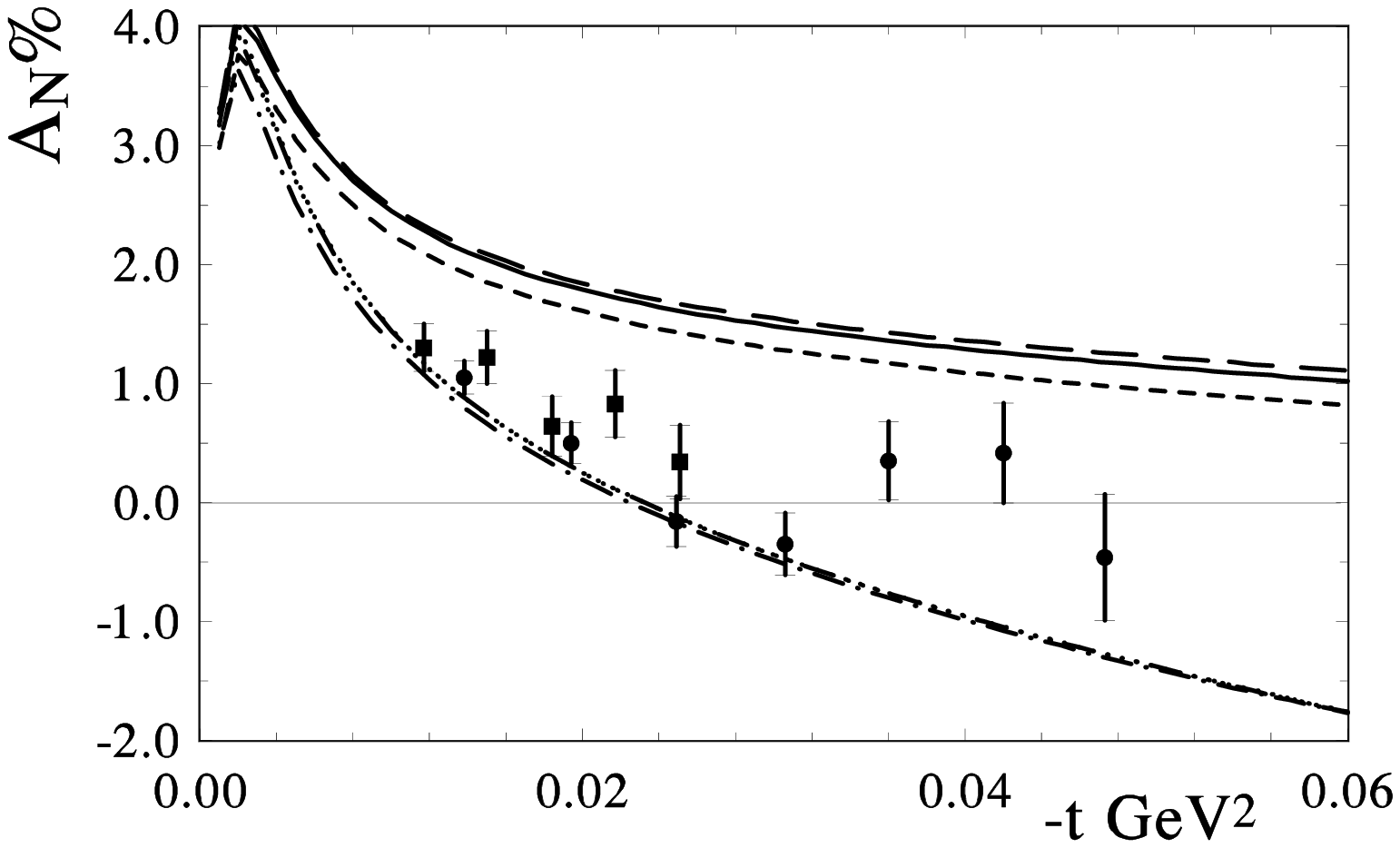,width=7.8cm,height=6cm}}
{\small{\bf Figure 2.}
 $A_N$ without ( upper 3 curves) and with  hadron-spin-flip amplitude in 
          case I ($B^{-} = B^{+}$) for $p_L = 24, \ 100, \   250 \  $GeV/$c$.
          (dash-dot, dashed-dots, and dots correspondingly).
}
%\medskip
\end{wrapfigure}
%To insert figure (with the help of wrapfig.sty)\\

%%%%%%%%%%%%%%%%%%%%%%%%%%%%%

 We take  $\rho^{pA} = \rho^{pp}/2$  as $a_2$ and $\rho$ contributions 
  decrease in the  nucleus.   
 It is  possible that in hadron 
   scattering the ratio of the spin-non-flip to the 
    asymptotic part of the spin-flip amplitude decreases
    very slowly with energy. In this case, if we take in our analysis only this
   part of the spin-flip amplitude, we cannot make its real part proportional
   to  $\rho_{pp}$ in this energy region.

For the determination of $ \A^{pA}_{nf}(s,t)$, we rely
on the data obtained by the SELEX Collaboration \cite{selex}.
We also will  consider the possibility of 
  normalising  $B^{+}$ on the experimental
  data of \cite{shiz}. 
We assume that the slope slowly rises with $\ln{s}$ in a way 
similar to the $pp$ case.

 According to the above analysis we investigate two  variants  for the  
 slope of the spin-flip amplitude: case I -
 $B^{-} \ = \ B^{+}$; case II - $B^{-} \ = \ 2 B^{+}$.
 The coefficients $k_1$ and $k_2$ are chosen to obtain the best description
of $A_N$ 
\ba
  A_N\frac{d\sigma}{dt} =
         - 4 \pi [Im(\A_{nf})Re(\A_{sf})-Re(\A_{nf})Im(\A_{sf})],
\ea 
at  $  p_L = 24, \ 100 $~GeV/$c$. 
Of course, we only aim at a qualitative description
as the data are only preliminary and as they are
normalised to those at $p_L = 22 $~GeV/$c$ \cite{an22}.

  %%%%%%%%%%%%%%%%%%%%%%%%%%%%%%%%%%%%%%%%%%%%%%%%%%
\begin{wrapfigure}{R}{8cm}
\mbox{\epsfig{figure=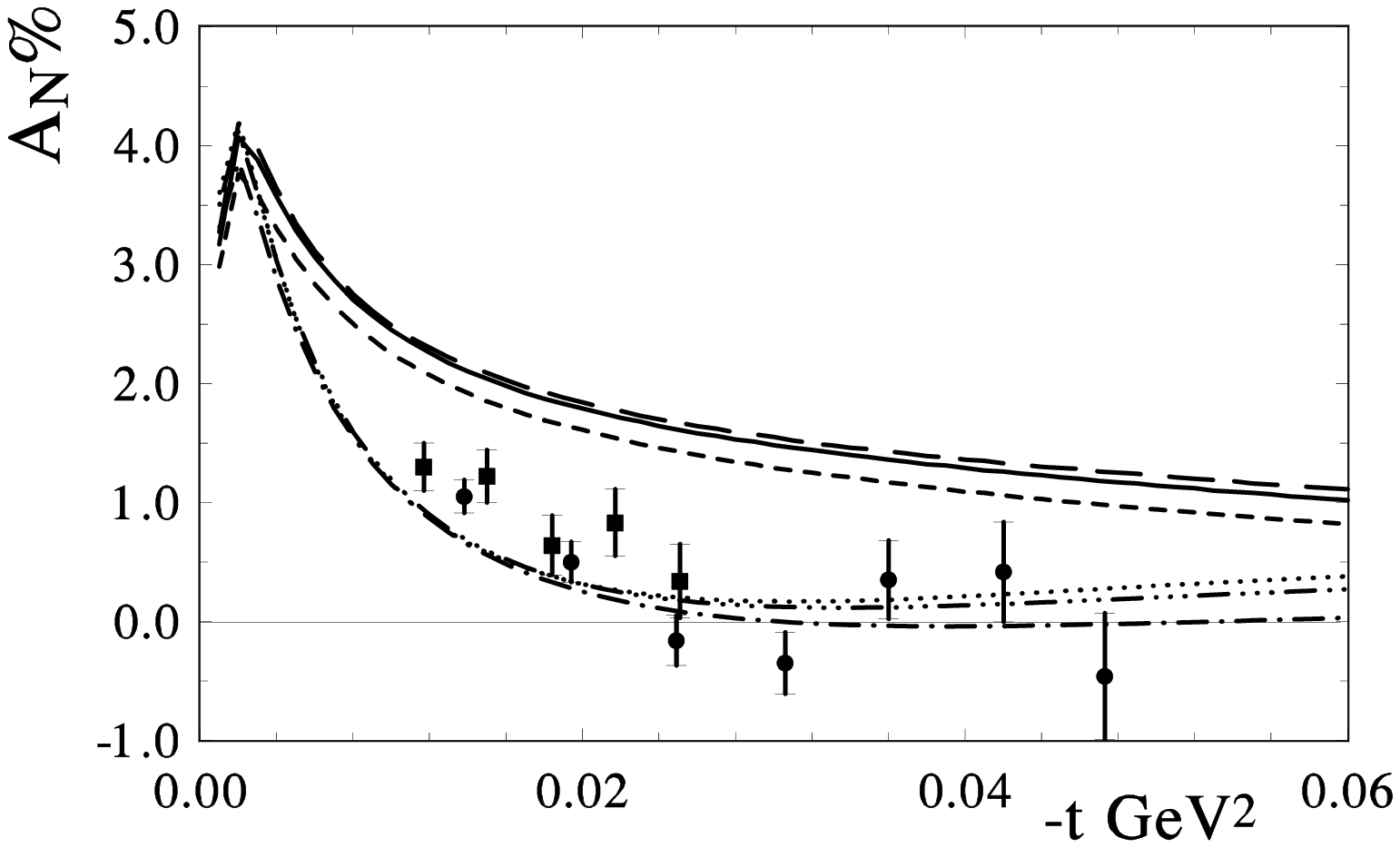,width=7.8cm,height=6cm}}
{\small{\bf Figure 3.}
$A_N$ with hadron-spin-flip amplitude in 
          case II ($B^{-} =2 B^{+}$) for 
      $ p_L = 24,  \ 100, \ 250 \ $GeV/$c$.
       (dash-dot, dash-dots, and dots correspondingly). 
}
%\medskip
\end{wrapfigure}
%To insert figure (with the help of wrapfig.sty)\\

%%%%%%%%%%%%%%%%%%%%%%%%%%%%%

   In Fig.2 and Fig. 3,
   the calculations are made at $p_{L} =  \ 100 \  $~GeV/$c$
  for the different normalisation of the slope $ B^{+}$ 
  (on  data of \cite{shiz}  and 
  \cite{selex}). At $p_{L} =  \ 100 \  $~GeV/$c$, they give 
  $B^{+} = 58.3 \ $GeV$^{-2}$ and  $B^{+} = 72.1 \ $GeV$^{-2}$ respectively.
  It is clear that this difference  changes the size of $A_N$ at
  $|t| \geq 0.02 \ $GeV$^2$  only slightly. 
  We can see that in both case we obtain a small energy dependence.
  In  case I, 
  the $t$- dependence of $A_N$  is weaker immediately
  after the maximum. But at large $|t| \geq 0.01 $GeV$^2$, the behaviour of
  $A_N$ is very different:  we obtain different signs for $A_N$ at
  $|t| \approx 0.06 \ $GeV$^2$.  
   In case I when $B^{-} = B^{+}$,  $A_N$ changes its sign 
  in the region  $|t| \approx 0.02$ and then grows in magnitude.
 
 In case II,   when $B^{-} = 2 B^{+}$,  $A_N$ approaches zero
   and then grows  positive again. 
  It is interesting to note that in more complex cases \cite{sc}, 
  where one investigates the analysing power for $p ^{12}C$-reaction 
  in case I, but with a more complicated form factor,
  one again  obtains the possibility that
   the slope  of the hadron spin-flip  exceeds the value
  $60\ $GeV$^{-2}$, and one can show that both slopes at very small momentum
  transfer are  equal to about  $90\ $GeV$^{-2}$.  
  Of course, such a large slope 
  for the spin non-flip amplitude requires  additional explanations 
  and cannot be obtained in the standard Glauber approach.

  We should note that all our consideration are based
  on the usual assumptions that
  the imaginary part of the high-energy scattering amplitude
  has an exponential behaviour.
  The other possibility, that  the slope changes slightly
   when  $ t \rightarrow 0 $, requires a more refined discussion
   that will be the subject of a subsequent paper.

\end{document}